\documentclass{iau307}
\usepackage{graphicx}
\usepackage{natbib}
\usepackage{url}
\usepackage{dtklogos}
\bibpunct{(}{)}{;}{a}{}{,}

\title[DACs in LBV binary MWC~314] 
{Discrete absorption components in the massive LBV Binary MWC 314}

\author[Lobel, Martayan, Corcoran, Groh, \& Fr\'{e}mat]   
{A. Lobel$^1$,
C. Martayan$^2$,
M. Corcoran$^3$,
J.H. Groh$^4$,
\and Y. Fr\'{e}mat$^1$
}

\affiliation{$^1$ Royal Observatory of Belgium, Brussels \\ email: {\tt Alex.Lobel@oma.be, alobel@sdf.org} 
\\[\affilskip]
$^2$European Southern Observatory, Chile
\\[\affilskip]
$^3$Goddard Space Flight Center, Greenbelt MD, USA
\\[\affilskip]
$^4$Geneva Observatory, Versoix, Switzerland
}

\pubyear{2014}
\volume{307} 
\pagerange{}
\jname{New windows on massive stars: asteroseismology, interferometry, and spectropolarimetry}
\editors{G. Meynet, C. Georgy, J.H. Groh \& Ph. Stee, eds.}

\begin{document}

\maketitle

\begin{abstract}
We investigate the physical properties of large-scale wind structures around massive 
hot stars with radiatively-driven winds. We observe Discrete Absorption Components 
(DACs) in optical He~{\sc i} P Cygni lines of the LBV binary MWC 314 ($P_{\rm orb}$=60.8 d). 
The DACs are observed during orbital phases when the primary 
is in front of the secondary star. They appear at wind velocities between $-$100 
$\rm km\,s^{-1}$ and $-$600 $\rm km\,s^{-1}$ in the P Cyg profiles 
of He~{\sc i} $\lambda$5875, $\lambda$6678, and $\lambda$4471, signaling high-temperature 
expanding wind regions of enhanced density and variable outflow velocity. 
The DACs can result from wave propagation linked to the orbital motion near the 
low-velocity wind base. The He~{\sc i} lines indicate DAC formation close to the 
primary's surface in high-temperature wind regions in front of its orbit, or in 
dynamical wind regions confined between the binary stars. We observed the DACs 
with Mercator-HERMES on 5 Sep 2009, 5 May 2012, and 
6 May 2014 when the primary is in front of the secondary star. 
XMM-Newton observations of 6 May 2014 significantly detected MWC 314 in X-rays at an average rate of 
$\sim$0.015~$\rm cts\,s^{-1}$.  

\end{abstract}

\firstsection 
\section{Introduction}
Current understanding of the role of the Luminous Blue Variable (LBV) stage in the evolution 
of the most massive stars is very limited. Important questions are how the eruptions of several
$\rm M_{\odot}$ are triggered, and what role binarity plays in the properties of the LBV 
stage and in shaping the nebulae observed around LBVs. For example, is most of the 
mass lost by an LBV star due to a steady radiatively-driven stellar wind, or is the 
H-rich envelope removed by punctuated eruption-driven mass loss? There are two 
confirmed LBV binaries: $\eta$ Car in the Galaxy, and HD~5980 in the SMC. 
The extremely luminous star MWC 314 has recently been recognized ($P_{\rm orb}$ = 60.8 d and $e$ = 0.23) as 
an eclipsing massive binary system \citep{2013AA...559A..16L}. Indications that MWC 314 is 
a (dormant) LBV are strong. 
The star is very luminous \citep*[log $L$/$\rm L_{\odot}$ = 5.9; see also][]{1998AAS..131..469M}
with an optical spectrum and SED nearly identical to the
canonical LBV P Cygni. MWC 314 is surrounded by a bipolar H$\alpha$ 
nebula that may be the result of an eruption of MWC 314 more than 
100,000 years ago \citep{2008AA...477..193M}. 
MWC 314 is a single-lined, eclipsing binary system with masses of 40$+$26 $\rm M_{\odot}$, and 
radii of 87$+$20 $\rm R_{\odot}$. The largest diameter of the binary system is $\sim$1.1 AU. 

\begin{figure}
\begin{center}
\includegraphics[width=6.cm, height=12cm, angle=270]{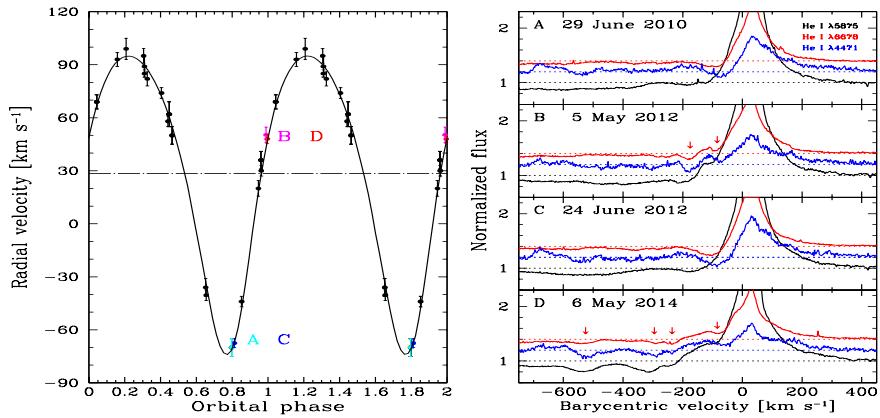} 
\caption{{\bf Left-hand panel}: orbital radial velocity curve of MWC 314. Two orbital eclips phases B and D when the primary is in front of the secondary ($\phi$$\sim$0), and two quadrature phases A and C ($\phi$$\sim$0.8) are marked. {\bf Right-hand panels}: the profiles of three He~{\sc i} lines are shown for orbital phases marked A, B, C, and D. DACs are observed in the violet wings of the lines ({\it marked with arrows}) during the orbital phases B and D. XMM-Newton observed X-rays on 6 May 2014 (D).
}
\label{fig1}
\end{center}
\end{figure}

\section{Radial velocity curve and DACs in He~{\sc i} P Cyg lines}
We measure the radial velocity curve in Fig. 1 from selected absorption 
lines of the primary star. Two spectra marked with A and C are observed in the 
quadrature phase close to minimum radial velocity of the primary 
(0.75 $<$ $\phi$ $<$ 0.85). Two spectra observed when the primary is in front of the 
secondary are marked with B and D (0.95 $<$ $\phi$ $<$ 1.05). Three He~{\sc i} lines reveal 
P Cyg profiles with wind expansion velocities to $\sim$1200 $\rm km\,s^{-1}$. 
On 5 May 2012 (B) we observe two discrete absorption components in the 
extended violet wings of the three lines. The DACs are observed at wind 
velocities of $\sim$100 $\rm km\,s^{-1}$ and $\sim$180 $\rm km\,s^{-1}$ 
in the line profiles ({\em marked with arrows}).  
The DACs are not clearly observed in the quadrature phases of 29 June 2010 (A) 
and 24 June 2012 (C). We observe four DACs on 6 May 2014 (D) at wind 
velocities between $\sim$100 $\rm km\,s^{-1}$ and $\sim$600 $\rm km\,s^{-1}$.

\section{Conclusions}
Our high-resolution spectroscopic monitoring program of MWC 314 
with Mercator-HERMES during the last 5 years reveals DACs in the orbital phases when 
the primary is in front of the secondary star on 5 Sep 2009, 5 May 2012, and 6 May 2014. 
The high-excitation temperatures of the He~{\sc i} 
lines signal expanding wind regions of enhanced density and variable outflow velocity. 
The DACs can form close to the primary's surface (of B0-type) in high-temperature and 
density-enhanced wind regions in front of its orbit. The recurrence of the DACs 
in orbital phases when the primary is in front of the secondary can result from wave 
propagation which is physically linked to the orbital motion near the (low-velocity $<$ 150 $\rm km\,s^{-1}$) 
wind base. Alternatively, the DACs can form in dynamical (high-temperature) wind regions 
confined between the binary stars at the shock interface of a colliding wind region. We expect 
orbital X-ray variability in MWC 314 (similar to the LBVs $\eta$ Car and HD~5980) during 
planned XMM-Newton observations in the quadrature phase of Oct. 2014.






\bibliographystyle{iau307}
\bibliography{MyBiblio}







\end{document}